\documentclass[final,5p,times,twocolumn,sort&compress]{elsarticle}

\usepackage{amssymb}
\usepackage{amsmath}
\usepackage{empheq}
\usepackage{comment}
\usepackage{balance}
\usepackage{placeins}
\usepackage{enumerate}
\usepackage[inline,shortlabels]{enumitem}
\usepackage[usenames,dvipsnames]{color}
\usepackage{xcolor}
\usepackage[colorlinks=true,linkcolor=Maroon,citecolor=Maroon,urlcolor=Maroon]{hyperref}
\usepackage[utf8]{inputenc}
\usepackage[english]{babel}
\usepackage{geometry}
\usepackage{booktabs}
\usepackage{array}
\usepackage{longtable}
\usepackage{graphicx}
\usepackage{fancyhdr}
\usepackage{titlesec}
\usepackage{enumitem}
\usepackage{hyperref}
\usepackage{caption}
\usepackage{float}
\usepackage{url}

\begin{document}

\begin{frontmatter}

\title{Harnessing the ``Reactive Falling Effect'' for rehabilitation and performance boosting}

  \author[add1]{Paul-Emmanuel Sornette}
 \author[add1,add2]{Didier Sornette\corref{cor1}}
 
  \address[add1]{\scriptsize
        Logic Workout GmbH
    }
 
 \address[add2]{\scriptsize
        Institute of Risk Analysis, Prediction and Management (Risks-X),
        Academy for Advanced Interdisciplinary Sciences,\\
        Southern University of Science and Technology, Shenzhen, China
    }
    \cortext[cor1]{Corresponding author. Email: dsornette@ethz.ch}
    
\begin{abstract}
We present Logic Workout, a novel training method based on radical dynamic instability combining rolling, deformation and spring instability on small fitballs.
By exploiting the ``reactive falling effect'', it re-engages innate balance mechanisms and forces real-time motor corrections, leading to better joint control, coordination, and movement precision. Preliminary results from a cohort of 18 participants show complete resolution of chronic pain, enhanced functional mobility, and surprising improvements in strength and performance  \textemdash challenging the belief that instability training impairs power output. We hypothesise that, by harnessing maximal instability, Logic Workout activates deep neuromuscular pathways and improve rehabilitation outcomes, training efficiency, and athletic performance.

\vskip 0.3cm
\noindent
{\bf Keywords}: Dynamic Instability Training, Reactive Falling Effect, Neuromuscular Control, Rehabilitation, Performance Enhancement, Logic Workout

\end{abstract}
\end{frontmatter}

\section{Introduction}

Chronic and recurrent musculoskeletal and joint pain \textemdash whether in the back, elbows, knees, or other regions \textemdash represents a significant global health challenge affecting individuals across all age groups \cite{Vosglobal2023,Dahlhamer2018,Ricard2023}. These conditions arise from a broad range of causes, including sports injuries, repetitive strain, aging-related degeneration, and pathological disorders such as osteoarthritis and tendonitis. According to global health data, musculoskeletal disorders are among the leading contributors to disability worldwide, with low back pain ranked as the single most disabling condition. In the United States alone, an estimated 20.9\% of adults (approximately 51.6 million people) live with chronic pain.
Importantly, the burden of musculoskeletal pain is not confined to adults: pediatric chronic pain also presents a major concern, with conservative estimates suggesting that 20\% to 35\% of children and adolescents experience chronic pain worldwide \cite{Friedrichsdorf16}. 
This pervasive and multifaceted burden highlights the urgent need for scalable, effective, and age-adaptable strategies for prevention, management, and rehabilitation of joint and musculoskeletal pain.

Amid the growing burden of musculoskeletal disorders and chronic joint pain, there is widespread and increasing recognition of the importance of regular physical activity for maintaining both physical and mental health. The World Health Organization (WHO) recommends that adults engage in at least 150 to 300 minutes of moderate-intensity aerobic activity per week to reduce the risk of cardiovascular disease, improve metabolic function, and critically, to prevent the decline of musculoskeletal health \cite{WorldHealth2020,Ambrose2015,Pedersen-Saltin2015}. Regular exercise has also been shown to reduce inflammation, increase joint lubrication, and enhance mobility  \textemdash all of which contribute to the prevention of joint-related pathologies and functional deterioration.

Beyond general health promotion, exercise is universally acknowledged as a cornerstone of both performance enhancement and therapeutic rehabilitation. It serves a diverse range of purposes: maintaining joint integrity, managing chronic pain, accelerating post-injury recovery, and supporting long-term physical function in both clinical and athletic populations \cite{Geneen2017,Hayden2021}. A growing body of evidence supports the role of targeted exercise interventions not only in symptom control, but also in addressing the underlying neuromuscular imbalances and coordination deficits that often accompany chronic pain syndromes and joint dysfunctions \cite{Ageberg2015,Lepley2017}.

To address the wide range of health, performance, and rehabilitation needs, a diverse array of exercise modalities has been developed and continually refined through decades of clinical practice and sports science. These methods span from foundational techniques such as classical gymnastics, calisthenics, and weight training, to more therapeutic or corrective approaches like Pilates, yoga, aquatic therapy, resistance band training, and proprioceptive neuromuscular facilitation (PNF). In recent years, more functional and innovative strategies have gained popularity, including TRX suspension training, kettlebell workouts, dynamic stretching, and core stabilization routines.

Alongside traditional exercise modalities, there has been a growing emphasis on training tools that incorporate elements of balance and instability \textemdash  such as Bosu balls, wobble boards, balance discs, foam rollers, and large Swiss balls \textemdash to enhance joint stability, proprioceptive awareness, and neuromuscular control. This development has led to increased interest in Instability Resistance Training (IRT), which deliberately employs unstable surfaces to elicit greater activation of deep stabilizer muscles, refine motor coordination, and improve postural control. Evidence from clinical studies and systematic reviews suggests that IRT can enhance core and spinal stabilizer engagement more effectively than equivalent exercises performed on stable surfaces.
 As a result, IRT has been increasingly recommended for integration into rehabilitation programs targeting conditions such as low back pain, anterior cruciate ligament (ACL) injuries, chronic ankle instability, and fall prevention in older adults \cite{Behm2013,Behm-Colado2012}.
However, it is important to recognize that instability training methods are often associated with trade-offs, including reduced maximal force output, power generation, and movement velocity \cite{BehmAnderson06}, which may make IRT less suitable as a primary method for developing strength or explosive power in athletic populations. 

Building on a growing body of research supporting instability training, we propose to extend this concept to its radical extreme through a novel set of exercises collectively referred to as the Logic Workout (LW), specifically designed to exploit maximally unstable surfaces  \textemdash namely, small fitballs (also known as mini stability balls). 

The LW  exercises are built around what we call the reactive falling effect: the highly efficient neuromuscular reflexes humans develop in early childhood  \textemdash particularly during the process of learning to walk  \textemdash which enable rapid, instinctive muscle activation to prevent falls. Logic Workout targets these reflexes by creating conditions that eliminate compensatory movement patterns, forcing full-body neuromuscular engagement and promoting deep, automatic stabilization responses. The reactive falling effect harnesses the unique geometry and elasticity of small fitballs  \textemdash  combining rolling (multi-directional movement without fixed pivot points), deformation (surface shape shifting under load, altering contact dynamics), and spring instability 
(oscillations produced by elastic rebound)  \textemdash  to enable an entirely new universe of movement and exercise possibilities.
The combined effects of unstable motion and deformation compel the nervous system to abandon habitual control 
strategies and engage in continuous, finely tuned micro-adjustments.
 This stimulates the full spectrum of muscle fibers, reveals hidden weaknesses, and facilitates automatic neuromuscular correction. Unlike 
 instability training methods, which often allow individuals to mask deficits with compensatory movements, LW makes such compensation impossible \textemdash leading to faster, deeper, and more permanent motor adaptations.

We present preliminary results from a cohort of 18 individuals who have been training with LW. 
Participants report enhanced joint control, reduced chronic pain, improved functional mobility, and, notably, unexpected gains in athletic performance.
These outcomes challenge the conventional belief that instability training inherently compromises maximal force output, power generation, or movement velocity. Even seemingly standard exercises, such as push-ups, take on entirely new dimensions under conditions of radical instability. The increased neuromuscular demands amplify training efficiency, turning familiar movements into powerful stimuli for strength, coordination, and motor control.

While further systematic studies are needed to rigorously validate these findings, our initial results support the hypothesis that maximal instability, when harnessed correctly, can significantly accelerate the development of joint resilience and neuromotor adaptability. By leveraging this radical form of instability training, Logic Workout offers the potential to redefine both therapeutic rehabilitation and high-performance training \textemdash tapping into the body's innate ability to stabilize and protect itself under stochastic real-world conditions.

\section{Logic Workout: Designing an Optimized Brain-Body Training System}

\subsection{Core Design Principles}

Developed by the first author through over a decade of trial and errors, Logic Workout (LW) is an innovative training system inspired by the principles of structural engineering \textemdash specifically, the dynamic redistribution of forces to maintain equilibrium. At its foundation lies Newton's Third Law: every action generates an equal and opposite reaction. In the human body, this principle is expressed through coordinated muscle activation to resolve reactive forces, particularly under unstable conditions.

Logic Workout integrates foundational elements from Pilates, yoga, martial arts, and traditional resistance training \textemdash such as planks, push-ups, and single-leg stances \textemdash but reengineers these familiar movements by introducing a reactive falling effect. This is achieved through the use of a small, free-rolling fitball, which eliminates fixed pivot points and external stabilizers. Unlike semi-stable tools like BOSU balls, TRX systems, rocker boards, or conventional free weights that allow users to anchor and stabilize through selective muscular engagement, the small fitball creates multi-axis, gravity-driven instability. This prevents reliance on compensatory strategies and channels reactive forces through the entire kinetic chain.

Unlike larger Swiss balls or semi-stable platforms like the Bosu, small fitballs, typically around 20 cm in diameter, offer minimal contact area and, when incorporated into carefully structured movements, generate a markedly high degree of dynamic instability. This challenges the body's balance and postural control systems in novel and demanding ways. Due to their small size, the rolling and deformation response time of these fitballs closely matches the human neuromotor reaction time, approximately 100 milliseconds. Even minor lateral displacements of just a few centimeters are sufficient to create significant angular shifts, optimally stimulating the neuromuscular control system and requiring continuous, high-precision adjustments for stabilization.

The rolling and deformable nature of the surface activates innate threat-detection mechanisms, shifting the brain's focus toward real-time stabilization of musculature, ligaments, and fascial networks. The perceived risk of falling forces both conscious and unconscious systems to engage fully, triggering deep neuromuscular responses. Subtle shifts in balance demand continuous adjustments in posture and joint alignment, transforming gravity and bodyweight into powerful tools for proprioceptive stimulation, neuromuscular activation, functional training, and rehabilitation.

\subsection{Closed Fists on Small Fitballs}

Except in beginner-level variations, a set of Logic Workout exercises are performed with closed fists placed on fitballs to enhance the rolling and deformation instabilities.

\subsubsection{Biomechanical Advantages of the Closed Fist Position}
 
 This configuration serves multiple purposes:
\begin{enumerate}
  \item \textbf{Increased Range of Motion:} Closed fists elevate the contact point by approximately 5 cm, increasing the range of motion and intensifying the stimulus per repetition.
  \item \textbf{Maximized Rolling Instability:} A closed fist minimizes the contact surface, allowing the fitball to roll freely and respond dynamically to pressure shifts.
  \item \textbf{Enhanced Deformation:} Concentrating load through the fists increases local compression, amplifying surface deformation and instability.
  \item \textbf{Wrist Stabilization:} Fist-based support inherently engages and strengthens the wrist joint, a common point of weakness.
\end{enumerate}

\subsubsection{Agonist-Antagonist Synergy}

Logic Workout employs three fist orientations \textemdash \textit{neutral}, \textit{pronation}, and \textit{supination} \textemdash to activate distinct muscle chains and replicate both pushing and pulling mechanics:
\begin{itemize}
  \item \textbf{Neutral:} Fists are parallel to the body (default configuration). Emphasizes pushing movements involving the chest, shoulders, and triceps.
  \item \textbf{Pronation:} Fists perpendicular, palms down. Simulates the eccentric phase of a pronated pull-up, engaging upper back, core, and arm musculature.
  \item \textbf{Supination:} Fists perpendicular, palms up. Mirrors the eccentric phase of a chin-up, emphasizing the biceps and back. This is typically the most challenging variation.
\end{itemize}

\subsection{Training Methodology}

\subsubsection{Standard LW protocol}

Logic Workout comprises over 45 distinct exercises, each available in up to five progressively challenging levels  \textemdash ranging from Entry to Master  \textemdash and encompassing movements for the upper body, core, and lower limbs.
Exercises are organized into structured sets of four movements per session.

A typical Logic Workout session includes:
\begin{itemize}
  \item A 3-minute dynamic warm-up to prepare the body.
  \item A 60-second HIIT (high intensity training interval) burst to elevate cardiovascular activity.
  \item Four consecutive Logic Workout exercises (90 seconds each), targeting specific muscle groups, performed without rest.
\end{itemize}

This core sequence lasts approximately 10 minutes. Depending on the user's capacity, the circuit may be repeated up to four times, with 90 seconds of rest between sets, for a total duration of up to 40 minutes. Empirical observations indicate that performance typically declines by the fourth set, suggesting synergistic physical and cognitive fatigue \textemdash a hallmark of the system's integrated neuromuscular demands.
We stress that, with Logic Workout's use of radical instability, standard exercises may become highly challenging, leading to superior gains in strength, coordination, and performance as reported below.

A selection of representative exercises from the Logic Workout program can be viewed at: \url{https://logicworkout.app.link/e/GC6LyKz3UTb}.

\subsubsection{Ladder Technique}

This technique introduces strategic pauses at multiple points during each movement. For example, during a fitball push-up, one may pause in the plank position at various heights for three seconds. These pauses enhance muscle engagement, coordination, and control \textemdash transforming simple movements into multidimensional challenges and accelerating adaptation.

\subsubsection{Kinesthetic Technique (Sensory Deprivation)}

This method involves performing exercises with eyes closed, eliminating visual feedback. Without visual input, the nervous system must rely solely on proprioception to maintain balance and control. This dramatically increases the demand on internal sensory pathways, improving responsiveness, coordination, and core stability. Though exercises may need to be simplified initially, this technique significantly enhances body awareness and reaction time.

\section{Empirical Evaluation of Logic Workout}

We now report a first study designed to evaluate the effectiveness of the Logic Workout method across a diverse range of demographics and training objectives. 

\subsection{Study design}

A total of eighteen participants were recruited through convenience sampling. They ranged in age from 14 to 67 years and represented a wide spectrum of training backgrounds, including elite Olympic athletes, world champions, former combat athletes, recreational trainees, sedentary individuals and patients undergoing rehabilitation. Their medical histories were equally varied, encompassing post-surgical recovery, chronic pain syndromes, athletic performance plateaus, adolescent joint pain, golfer's elbow, and knee injuries.

The intervention protocol consisted of Logic Workout sessions incorporating dynamic instability training on small fitballs, with exercises progressing in difficulty over time. Table \ref{htbwgrb} summarises the main intervention parameters.

\begin{table}[H]
\centering
\caption{Intervention Parameters}
\begin{tabular}{@{}ll@{}}
\toprule
\textbf{Parameter} & \textbf{Range} \\
\midrule
Frequency & 1-7 sessions per week \\
Duration & 3-40 minutes per session \\
Total weekly volume & 3 minutes to 3 hours \\
Training period & 2 weeks to 6 months \\
Technique & Logic Workout exercises \\
\bottomrule
\end{tabular}
\label{htbwgrb} 
\end{table}

Exercises were generally performed either at the participants' homes or in gym settings, with regular weekly interviews conducted to monitor progress, provide guidance, and offer corrective advice when necessary to ensure proper execution of the exercises.

\subsection{Outcome Measures}

The measured primary outcomes include strength performance (such as repetition maximums and load capacity), subjective reports of pain and functional limitations, and training efficiency, defined as results achieved relative to time invested. 

Secondary outcomes encompassed cardiovascular improvements, changes in functional capacity, injury prevention and recovery, and body composition changes. 

Data were collected through self-reported assessments, with participants providing baseline and post-intervention metrics related to performance, pain, and overall function.

\section{Results}

Eighteen participants were evaluated, categorized into two primary groups: those undergoing rehabilitation or seeking pain relief, and those aiming for performance enhancement. All individuals in the rehabilitation group had previously attempted other treatment methods without achieving satisfactory results. Participants in the performance group were already physically active, with moderate to advanced training backgrounds. 
As described below, the results demonstrate consistent effectiveness of the Logic Workout approach across diverse populations and objectives, highlighting its versatility in both therapeutic and performance-oriented contexts.

\subsection{Rehabilitation and Pain Relief Cases}

\subsubsection{Florian, 40 (Male, Chronic Back Pain)}
\textbf{Background:} Five-year history of severe L4-L5 back pain. Previous unsuccessful treatments included manipulations and physiotherapy. Scheduled for surgical intervention.

\textbf{Intervention:} Initial phase: 20-minute sessions (1h20/week) progressing to 30-minute sessions (2h/week) post-recovery. \textbf{Duration:} 6 months.

\textbf{Results:} Complete pain resolution within 3 months, Surgery cancellation, Master-level exercise achievement at 4 months (10 fist push-ups on single fitball)

\subsubsection{Didier, 67 (Male, Post-Surgical Recovery)}
\textbf{Background:} Post-surgical recovery following an open olecranon fracture and an upper humerus fracture, treated with titanium rod implantation in the humerus and cable-screw fixation in the olecranon. Limited recovery prognosis due to age and injury severity.

\textbf{Intervention:} Daily 10-minute rehabilitation protocol. \textbf{Duration:} 4 months.

\textbf{Results:} Restoration of strength and mobility, successful return to extreme sports such as kitesurfing and snowboard carving, and consistent performance of 60 push-ups within a high-intensity interval training (HIIT) format.

\subsubsection{Nicholas, 14 (Male, Adolescent Athlete)}
\textbf{Background:} Competitive fencer with one-year chronic knee and hip pain. Multiple failed conventional treatments led to competition withdrawal.

\textbf{Intervention:} Acute phase: 4 specific exercises, 10 min/day for 1 week. Maintenance: standard program 2.5h/week. \textbf{Duration:} 1 week acute + 1 month maintenance.

\textbf{Results:} Complete pain resolution within one week, return to competition after one-year hiatus, enhanced strength and speed, won subsequent two competitions at the time of writing (June 4, 2025).

\subsubsection{Dan, 40 (Male, Golfer's Elbow)}
\textbf{Background:} Experienced weight trainer with golfer's elbow requiring one-year training cessation. 10kg excess body fat at baseline.

\textbf{Intervention:} Progressive protocol: 20-minute sessions (2h/week) advancing to 40-minute sessions (2h40/week). \textbf{Duration:} 4 months.

\textbf{Results:} Complete pain resolution at 2 months, 10kg fat loss with concurrent muscle hypertrophy, return to enhanced strength levels exceeding pre-injury capacity.

\subsubsection{Daniele, 40 (Male, Complex Knee Injury)}
\textbf{Background:} Former Muay Thai fighter with ACL rupture and tibial plateau fracture. Post-surgical complications included lateral collateral ligament damage and fibular head inflammation. Failed conventional rehabilitation attempts.

\textbf{Intervention:} Exclusive Logic Workout: 20-40 minute progressive sessions. \textbf{Duration:} 4 months (2.5h/week).

\textbf{Results:} 95\% knee flexion restoration (heel-to-glute contact), complete functional recovery where conventional methods failed,  enhanced strength exceeding pre-injury levels.

\subsubsection{Anthonia, 38 (Female, Advanced Pilates Practitioner with Chronic Lower Back Pain)}
\textbf{Background:} 1.5-year lower back pain history. Daily Pilates practitioner (30-40 min, 6 days/week) with minimal improvement.

\textbf{Intervention:} Replaced Pilates with Logic Workout (20 min/day). \textbf{Duration:} 3 months.

\textbf{Results:} Complete pain elimination within 2 hours total training time, superior strength gains compared to previous Pilates regimen,  enhanced daily energy and focus.

\subsubsection{Isabelle, 32 (Female, Neck and Shoulder Tension)}
\textbf{Background:} Daily yoga practitioner (20 min/morning) with persistent shoulder and neck tension.

\textbf{Intervention:} Two 3-minute Logic Workout exercises. \textbf{Duration:} 1.5 months.

\textbf{Results:} Complete pain resolution in 6 minutes total training time, no pain recurrence, notable strength increases.

\subsection{Performance Enhancement Cases}

\subsubsection{David , 28 (Male, Experienced Weight Trainer)}
\textbf{Background:} Experienced athlete training 6 hours weekly.

\textbf{Intervention:} Replaced 1 hour of weight training with Logic Workout (2×30min/week). \textbf{Duration:} 2.5 weeks (5 sessions), continued for 3 months.

\textbf{Results:} Squat: +25 lbs (250→275 lbs), improved repetitions (1→2 reps);  Incline bench: Personal best 10 reps at 60 lbs;
Dips: +30\% improvement (10→13 clean reps);  Enhanced stamina across all exercises; Continued personal records throughout 3-month period.

\subsubsection{Brendan, 38 (Male, Ex-Jujitsu Fighter)}
\textbf{Background:} Athlete-level fitness with performance plateau.

\textbf{Intervention:} Daily 10-minute sessions (fitball push-ups focus). \textbf{Duration:} 2 weeks.

\textbf{Results:} Progression from kneeling to full fitball push-ups; 30\% increase in maximum push-up repetitions beyond previous personal record.

\subsubsection{Scott, 31 (Male, Advanced Amateur)}
\textbf{Background:} Weight training 5 hours weekly.

\textbf{Intervention:} Modified routine: one 40-minute Logic Workout + 4 hours weight training weekly. \textbf{Duration:} 2 weeks (2 sessions total).

\textbf{Results:} 15\% improvement in both bench press and squat personal records.

\subsubsection{Additional Performance Cases}

\textbf{Matteo, 41 (Male, Sedentary Recovery):} After 6-month sedentary period, 2 months of Logic Workout (3×30min/week) resulted in ``excellent cardio'' upon running resumption, progression from inability to hold neck during abdominal exercises to 10 push-ups, and visible muscle hypertrophy.

\textbf{Constantin, 34 (Male, Recreational Trainer):} 2 months of Logic Workout (3×20min/week) produced bench press increase from 60kg to 85kg with fat loss and visible hypertrophy.

\textbf{Yury, 29 (Male, Racing Pilot):} 2 months of training increased push-up capacity from 5 to 15 repetitions with fat loss and visible hypertrophy.

\textbf{Jaufray, 29 (Male, Recreational Tennis Player):} Daily 3-minute sessions for 3 months resulted in 5kg fat loss, complete bilateral arm strength rebalancing, and knee pain resolution.

\textbf{Nathalie, 39 (Female, Elite Olympic Athlete):} 2019 World Champion fencer recovering from tumour surgery in the lumbar region, 3 months of comprehensive Logic Workout (2h/week) produced subjective assessment of being ``stronger than ever in career'' with notable speed increases and first-time achievement of 10 consecutive standard push-ups.

\textbf{Masha, 27 (Female, Postpartum Recovery):} After minimal results from 4-month Pilates program, 2.5 months of Logic Workout (3×30min/week) resulted in ``best shape of her life'' with complete fat loss and notable glute hypertrophy.

\textbf{Cecile, 30 (Female, Postpartum Recovery):} After one year of Pilates with minimal improvement, 2 months of daily 10-minute Logic Workout produced notable strength increase and energy enhancement with fat loss and muscle definition.

\textbf{Kaajal, 25 (Female, Recreational Swimmer):} Adding Logic Workout (3×20min/week) to existing swimming routine for 2 months resulted in first-time achievement of 2 push-ups with general strength increase and fat loss.

\section{Discussion}

\subsection{Key Findings}

One of the most striking outcomes of this study was the superior training efficiency demonstrated by Logic Workout. Participants achieved comparable or superior physical results with only 1-3 hours of weekly training, as opposed to the 5-6 hours typically required by conventional exercise programs. This suggests a 3- to 5-fold improvement in training efficiency, highlighting Logic Workout's potential for time-constrained individuals.

Another key finding was the rapid resolution of chronic pain conditions, which significantly outpaced conventional rehabilitation timelines. For example, one case of shoulder tension was resolved in just 6 minutes, compared to the multiple weeks usually required with standard therapy. Lower back pain was eliminated after only a total of 2 hours of training, whereas similar results with daily Pilates typically require several months. Golfer's elbow, a condition that typically requires 6 to 12 months for recovery, showed complete resolution within 2.5 months following the implementation of Logic Workout  \textemdash after a full year of unsuccessful attempts with other interventions, including BOSU training, resistance bands, localized massage, and cryotherapy.
A complex knee injury that had not responded to traditional rehabilitation was fully resolved in 4 months. Furthermore, a participant undergoing post-surgical recovery achieved full functional restoration despite a poor initial prognosis, underscoring the method's rehabilitative potential.

Logic Workout was also found to enhance performance in conventional training modalities. Participants consistently reported setting new personal records and experiencing improved stamina across a broad range of traditional exercises, indicating that Logic Workout functions effectively not only as a standalone system but also as a powerful cross-training complement. Its use of radical instability transforms standard exercises into highly demanding neuromuscular tasks, which likely accounts for the observed performance gains. By dramatically increasing coordination demands and muscle engagement, Logic Workout appears to surpass conventional training methods in both efficiency and overall training impact.

Importantly, the benefits of Logic Workout were observed across a broad age range, from 14 to 67 years. This age-independent effectiveness suggests that the dynamic instability principles underlying Logic Workout are universally applicable and adaptable throughout the human lifespan.

A particularly noteworthy case involved the rehabilitation of a world champion athlete following a tumour removal surgery. Not only did the athlete regain full strength, but she also reported performance levels exceeding their pre-treatment career peak. This suggests that Logic Workout holds promise for high-level athletic recovery, even in the face of severe physiological setbacks.

Finally, several participants exhibited improvements in body composition, including simultaneous fat loss and muscle gain. This indicates that Logic Workout may produce metabolic benefits beyond those typically observed in traditional strength training, potentially offering a dual advantage in both performance enhancement and aesthetic outcomes.

\subsection{Mechanisms of Action}

The effectiveness of Logic Workout appears to stem from the integration of biomechanical, neurological, and metabolic mechanisms that together support comprehensive physical and functional development.

A central principle of the system is its ability to disrupt compensatory movement patterns. By introducing radical dynamic instability, Logic Workout forces authentic neuromuscular engagement, revealing and correcting hidden weaknesses that are often masked in stable training environments. This mechanism ensures that the body's weakest links are exposed first, receiving the greatest training stimulus and thereby promoting musculoskeletal balance and injury prevention.

Simultaneously, Logic Workout promotes comprehensive muscle activation. Unlike traditional exercises that often isolate muscle groups, LW requires the coordinated effort of prime movers, stabilizers, and proprioceptive systems. This full-spectrum recruitment enhances both strength and coordination in a deeply integrated manner, contributing to functional improvements that transfer beyond the training context.

We propose that Logic Workout also facilitates neurological recalibration. The use of controlled instability may reactivate motor learning pathways first developed in early childhood \textemdash particularly those responsible for maintaining balance and reacting to perceived falling. This stimulation of dormant reflex circuits may accelerate neuromuscular adaptation and motor control, making the brain a more efficient regulator of movement.

Moreover, Logic Workout may offer distinct metabolic advantages. Training on unstable surfaces increases overall effort, caloric expenditure, and the activation of supporting musculature, which can enhance muscle protein synthesis, promote fat loss, and contribute to improved body composition.

Over time, the brain learns to optimize movement efficiency by integrating continuous sensory feedback with motor output, paralleling the adaptive load-balancing strategies observed in engineered tension-compression structures. We hypothesize that the ongoing feedback loop reinforces correct movement patterns while continuously adapting to instability, offering both immediate and long-term benefits for coordination, strength, and joint resilience.

\subsection{Clinical Implications}

These findings suggest that Logic Workout holds substantial promise across a range of clinical and performance contexts. It may serve as a primary training method for general strength development and functional conditioning. Additionally, it shows strong potential as a rehabilitation protocol for injury recovery and chronic pain management, particularly in cases where conventional methods have failed.

Furthermore, Logic Workout could be employed as a supplementary training tool to enhance the effectiveness of traditional exercise regimens. 
It may also function effectively as a maintenance program, supporting long-term health, mobility, and physical resilience.

Lastly, its impact on body composition underscores its potential as a valuable tool for simultaneously promoting fat loss and muscle gain, suggesting that Logic Workout may offer a comprehensive solution bridging both therapeutic needs and performance-oriented goals.

\subsection{Limitations}

This study has several limitations that should be acknowledged. First, the reliance on self-reported outcomes may introduce subjectivity and potential bias in the assessment of effectiveness. Second, the intervention protocols varied among participants, limiting the degree of standardization across the cohort. Third, the absence of a control group prevents direct comparisons with conventional training or rehabilitation methods. Finally, the current dataset lacks long-term follow-up, making it difficult to evaluate the durability and sustained impact of the observed benefits over time.
Preliminary in nature, this study is intended primarily to introduce the Logic Workout concept and motivate future, more rigorous empirical investigations to formally assess its efficacy across diverse populations and settings.

\section{A Neurobiological Hypothesis: The Reactive Falling Effect}

Human motor development and neuroplasticity are deeply shaped by the intrinsic dynamical instability of bipedalism. Unlike quadrupeds, humans must maintain upright posture while functioning as an unstable inverted pendulum, characterized by a high center of mass and a narrow base of support. This biomechanical challenge has played a defining role in the evolution of the human brain, which developed sophisticated systems for real-time sensory integration and motor control. Maintaining balance requires the continuous coordination of visual, vestibular, and proprioceptive inputs with motor outputs, a process mediated by neural circuits within the cerebellum, basal ganglia, and motor cortex \cite{Massion1994,Takakusaki2017} .

The idea that the brain evolved primarily to manage this instability offers a powerful framework for understanding both early development and adult learning. During infancy and early childhood \textemdash widely recognized as a critical period of heightened neural plasticity \textemdash the brain faces the monumental task of mastering postural control and locomotion. Through trial-and-error, environmental feedback, and intrinsic motivation, infants learn to stand, walk, and manipulate objects in space. These experiences shape neural architecture via mechanisms such as synaptic pruning and myelination, enabling the brain to optimize sensorimotor networks. Crucially, this process also supports the development of higher cognitive functions including language, social behavior, and problem-solving, suggesting a deep and synergistic relationship between motor and cognitive development \cite{Diamond2000,Thelen2000,AdolphHoch19}.

This perspective aligns closely with the work of neurobiologist Daniel Wolpert, who proposes that the brain's primary evolutionary function is to produce adaptable and complex movement \cite{Wolpert1,Wolpert2,Wolpert3}. As he succinctly puts it, ``We have a brain for one reason and one reason only: to produce adaptable and complex movement.'' Supporting this view, Wolpert highlights organisms like the sea squirt \textemdash an animal that digests its nervous system after attaching to a surface and ceasing to move \textemdash as evidence that complex nervous systems evolve in service of movement. In humans, his research shows how the brain constructs and refines internal models of motor commands, enabling it to predict outcomes, minimize variability, and rapidly adapt to changing environments. From this perspective, higher-order cognitive functions such as perception, memory, and decision-making are not isolated capabilities but evolved mechanisms that serve the broader goal of optimizing behavior through movement.

Taken together, these insights point to a fundamental conclusion: movement is not just an output of brain function \textemdash it is a core driver of brain development, organization, and adaptability. This raises a central question: can adults re-engage the brain's early-state learning capacity by reintroducing the same kinds of instability that shaped early development?

We propose that they can through targeted exposure to dynamical instability, delivered via a system of precisely designed exercises that place the brain and body into a heightened state of engagement. This is the foundation of Logic Workout, which induces a controlled reactive falling effect through movements performed on small, free-rolling fitballs. These exercises reactivate the brain's innate threat-detection systems and force continuous real-time corrections in posture, joint alignment, and core stabilization. In this environment, habitual and compensatory strategies are bypassed, and the nervous system is pushed into a state of deep sensorimotor activation. The result is a training experience that closely mirrors the neurophysiological conditions of early motor learning \textemdash facilitating the reactivation of dormant reflex pathways and the restoration of movement precision and control.

A growing body of research supports the view that successful motor training arises from the interaction of four key systems: (i) the central nervous system (CNS), including both spinal and cortical reflex loops \cite{Kandel2021}; (ii) peripheral mechanoreceptors that provide real-time sensory feedback \cite{ProskeGandevia12}; (iii) endocrine signals released by contracting muscles that modulate systemic responses \cite{Bostrom2012,Cotman2002,PedersenFeb2008,Pedersen2008}; and (iv) the muscles themselves, acting as both actuators and sensors \cite{Shadmehr2008}. We hypothesize that Logic Workout simultaneously stimulates all four components, creating a comprehensive and biologically grounded platform for physical training, rehabilitation, and performance enhancement.

Furthermore, we hypothesise that the radical instability and reactive falling effect at the core of Logic Workout re-engage sensorimotor integration processes similar to those that occur during early childhood \textemdash when walking, language acquisition, and spatial cognition co-develop. As such, the method may not only improve physical outcomes such as strength, balance, and coordination, but also promote cognitive adaptability and neuroplasticity, effectively reopening a developmental-like window through structured, body-centered learning.

\section{Conclusions}

Logic Workout dynamic instability training demonstrates preliminary evidence of exceptional effectiveness across diverse populations and training goals. The method appears to offer superior time efficiency, rapid pain resolution, and performance enhancement compared to conventional approaches. The universality of positive outcomes across age groups and fitness levels suggests broad clinical and practical applications.

The cases of complete recovery from chronic conditions that failed to respond to conventional treatments (golfer's elbow, complex knee injury) are particularly noteworthy and warrant further investigation through controlled studies.

Future research should include randomized controlled trials with standardized protocols, objective outcome measures, and long-term follow-up to confirm these preliminary findings and establish evidence-based implementation guidelines.

\vskip 0.5cm
{\bf Ackowledgements}: D.S. was partially supported by the National Natural Science Foundation of China (Grant No. T2350710802 and No. U2039202), Shenzhen Science and Technology Innovation Commission Project (Grants No. GJHZ20210705141805017 and No. K23405006), and the Center for Computational Science and Engineering at Southern University of Science and Technology.

\end{document}